\documentclass[a4paper]{PoS}

\title{Multi-messenger Astronomy: a Bayesian approach}

\ShortTitle{Multi-messenger Astronomy: a Bayesian approach}

\author{\speaker{G. Torralba Elipe}, R. A. Vazquez and E. Zas
  \\ Departamento de F\'{i}sica de
  Part\'{i}culas \& Instituto Galego de F\'{i}sica de Altas
  Enerx\'{i}as, Universidade de Santiago de Compostela, 15782,
  Santiago de Compostela, Spain\\ E-mail:
  \email{guillermo.torralba@usc.es},
  \email{vazquez@fpaxp1.usc.es} and 
  \email{zas@fpaxp1.usc.es }}

\abstract{After the discovery of the gravitational waves and the
  observation of neutrinos of cosmic origin, we have entered a new and
  exciting era where cosmic rays, neutrinos, photons and gravitational
  waves will be used simultaneously to study the highest energy
  phenomena in the Universe.  Here we present a fully Bayesian
  approach to the challenge of combining and comparing the wealth of
  measurements from existing and upcoming experimental facilities.  We
  discuss the procedure from a theoretical point of view and using
  simulations, we also demonstrate the feasibility of the method by
  incorporating the use of information provided by different
  theoretical models and different experimental measurements.}

\FullConference{35th International Cosmic Ray Conference --- ICRC2017\\
		10--20 July, 2017\\
		Bexco, Busan, Korea}

\usepackage{amsmath}
\usepackage{amsfonts}
\usepackage{amssymb}
\usepackage{caption}
\usepackage{float}
\usepackage{subcaption}

\newcommand{\sspace}{\mathcal{S}}

\newcommand{\mean}[1]{
\left \langle #1 \right \rangle
}

\newcommand{\ex}{$\text{E}_\text{X}$ }
\newcommand{\ey}{$\text{E}_\text{Y}$ }

\begin{document}

\section{Introduction}
We are incoming in a new era for astroparticle physics. A lot of
experiments are living together measuring different observables in a
wide range of energies: Pierre Auger Observatory \cite{auger},
Telescope Array \cite{ta}, HAWC \cite{hawc}, IceCube \cite{icecube},
Magic \cite{magic}, Antares \cite{antares}. New experiments will be
developed such as Cherenkov Telescope Array \cite{cta} or KM3Net
\cite{km3net} and we are going to have unprecedented number of events
to perform analyses that could answer the questions related with
high-energy cosmic rays, neutrinos and photons from more than century
ago such that: what are the cosmic rays?; where are they comming
from?; how are they accelerated?.  There are no doubts that for answer
these questions the experiments must share their results and the
answer will arrive by combining all the measurements.

In this work we present a brief review of Bayesian inference in
Sec. \ref{bayesian-inference}, explaining the parameter estimation and
hypothesis testing. The practice of these methods are shown using toy
simulations in Sec. \ref{simulations}. First we consider that two
experiments analyse different data without taking into account the
results of the other experiment. After that, we consider that the
experiments share their results and modify their prior information in
their analyses.Finally the conclusions are reported in
Sec. \ref{conclusions}.

\section{Review of Bayesian statistical inference}
\label{bayesian-inference}
The well known Bayes' theorem is a consequence of the law
of conditional probability
\begin{equation}
P(A|B,I) = \frac{P(A,B|I)}{P(B|I)},
\label{conditional_probability}
\end{equation}
and the law of total probability
\begin{equation}
P(B|I) = \sum_{i=1}^n P(B|C_i,I)P(C_i|I).
\label{total_probability}
\end{equation}
Here, $A$ and $B$ are two events of the {\it sample space}, $\sspace$
(the space of all possible outcomes of an experiment), the set
$\left\{C_i\right\}_{i=1}^{n}$ performs a partition of $\sspace$ and
$I$ is any {\it prior} information that we have before the analysis
(see Sec. \ref{parameter-inference}). The equation
\ref{conditional_probability} can be rewrite as
\begin{equation}
P(A,B|I) = P(A|B,I)P(B|I),
\label{conditional_probability2}
\end{equation}
which it is understood as: by assuming the information $I$ (which
include the prescription of probabilities), the probability of the
events $A$ and $B$ is the product of the probability of $A$ given $B$
(the probability of $A$ if $B$ occurs) and the probability that $B$
occurs. On the other hand \ref{total_probability} is readed as:
assuming $I$, the probability of $B$ is given by the sum of all
possibilities of obtaining $B$. Notice that since
$\left\{C_i\right\}_{i=1}^{n}$ is a partition of $\sspace$, either $B
= C_i$ for some $i$ or $B = \cap_{j=1}^k C_j$ for some $j$ and $k$.

Finally, the Bayes' theorem is expressed as 
\begin{equation}
\label{bayes_theorem}
P(A_j|B,I) = \frac{P(B|A_j,I)P(A_j|I)}{P(B|I)} = \frac{P(B|A_j,I)P(A_j,I)}{\sum_{i=1}^n P(B|A_i|I)P(A_i|I)},
\end{equation}
understood as: the probability of obtaining the event $A_i$ given $B$
and assuming $I$ is the product of the probability of obtaining $B$
given $A_i$ and the probability of obtaining $A_i$ normalised to all
possibilities of obtaining $B$.

\subsection{Parameter inference}
\label{parameter-inference}
Let $D=\{x_i\}_{i=1}^n$ be $n$ realizations of a random variable $X$,
{\it i.e,} $n$ results of experiments consisting in measuring the
variable $X$. Let $\theta$ be a parameter of interest.  Notice that
there are not restrictions on the dimensions of $X$ and $\theta$. The
Bayesian inference consists of allocating probabilities to the
possible values of $\theta$ according to the observed data set $D$
by solving the equation
\begin{equation}
\pi(\theta|D,I) = \frac{f(D|\theta,I)\pi(\theta|I)}{f(D|I)} = 
\frac{\text{Likelihood } \times \text{Prior}}{\text{Evidence}},
\label{eq:bayesian_inference}
\end{equation}
which is expressed in terms of probability density functions. Now we describe each term
appearing in Eq. \ref{eq:bayesian_inference}.
\\

{\bf Likelihood function: $f(D|\theta,I)$ } 

The likelihood function $f(D|\theta,I)$ is the conditional probability
distribution of $D$ given the unknown parameter $\theta$ and it is
usually denoted as $\mathcal{L}(\theta|D)$.  This function describes
how the data set $D$ is distributed assuming a given value of
$\theta$.  The likelihood function expresses all information
obtainable for the data satisfaying the {\it Likelihood principle}:
All the information about $\theta$ that can be obtained from an
experiment is contained in the likelihood function for $\theta$ given
$X$. Two likelihood functions for $\theta$ (from the same or different
experiments) contain the same information about $\theta$ if they are
proportional to one another, see \cite{berger_wolpert_book} and
\cite{birnbaum}. In \cite{birnbaum} it is also shown that the
likelihood principle is derived by the assumption of two principles:
the {\it principle of sufficiency} and the {\it principle of
  conditionality}.  These principles can be described informally as
asserting the ``irrelevance of observations independent of a
sufficient statistic'' (sufficiency) and the ``irrelevance of
experiments not actually performed'' (conditionality).  \\

{\bf The prior: $\pi(\theta|I)$} 

It describes all the information that we
have about the parameter of interest before performing the
experiment. A prior distribution can be created using information
about past experiments, using theoretical knowledge or expressing our
total ignorance about the problem. When we do not have information
about the parameter of interest one should follow the {\it Laplace
  criterion rule} paraphrased as: ``in the abscence of any further information (prior
information) all possible results should be considered equally
probable''. This kind of prior is the so called ``flat prior''.
\\

{\bf The posterior: $\pi(\theta|D,I)$} 

This function describes our knowledge about the $\theta$ parameter
after the data analysis of the experimental results. Then one can read Eq.
\ref{eq:bayesian_inference} as an update of the prior knowledge of
$\theta$, described by the prior, through the experiment described by
the likelihood. For each event $x_i \in D$ of the data set, our
knowledge about $\theta$ changes.  Once the posterior distribution is
known there are two standard estimators for the true value of
$\theta$: the mean of the posterior and the mode (the so called {\bf
  M}aximum of {\bf A P}osteriori distribution, MAP).
\\

{\bf The evidence: $f(D|I)$} 

Also denoted as $Z$ acts as a normalization constant in the parameter
inference but takes an important role in the {\it Bayesian Model Selection}
explained in Sec. \ref{bayesian-model-selection}. The evidence is given by:
\begin{equation}
Z = \int f(D|\theta,I)\pi(\theta|I) d\theta.
\end{equation}
\\
\subsection{Confidence intervals}
The confidence intervals or credible sets (here denoted as C.I) are
easy to calculate in the Bayesian approach. Once the posterior
distribution is known we want to find between which values
$[\theta_1,\theta_2]$ the actual value of the parameter has been
estimated.  Usually this question is answered with an associated
probability $q$ which is typically 0.68, 0.9 and 0.95. The limits of
the range are given by solving the equation
\begin{equation}
\label{credible_set}
q = P(\theta_{low} \le \theta \le \theta_{up}) = \int_{\theta_{low}}^{\theta_{up}} \pi(\theta|D,I) d\theta.
\end{equation} 
When the inferred value of $\theta$ equal or near to
one of the limits of the possible values of $\theta$, one
talk about upper or lower limits depending if $\theta \approx \theta_{min}$
or $\theta \approx \theta_{max}$.
\subsection{Bayesian model selection}
\label{bayesian-model-selection}
Consider now two hypotheses $I_1$ and $I_2$ that we want
to constrast and we perform an experiment which gives us the data set
$D = \{x_i\}_{i=1}^n$. We are going to consider that the likelihood
functions are different for the different hypotheses, for $I_1$ we
have $\mathcal{L}_1(\theta|D) = f_1(D|\theta)$ and for $I_2$ we have
$\mathcal{L}_2(\omega|D) = f_2(D|\omega)$ where $\theta$ and $\omega$
could in principle have different dimensions ($\theta$ could be for
instance a shape of an exponential distribution and $\omega$ could be
the mean and the variance of a normal distribution). The posterior
distributions are given by
\begin{equation}
\pi(\theta|D,I_1) = \frac{\mathcal{L}_1(\theta|D)\pi(\theta|I_1)}{Z_1}
\end{equation}
for the first hypothesis and 
\begin{equation}
\pi(\omega|D,I_2) = \frac{\mathcal{L}_2(\omega|D)\pi(\omega|I_2)}{Z_2}
\end{equation}
for the second hypothesis. $Z_1$ and $Z_2$ are the normalization
factors for their respective equations:
\begin{equation}
Z_1 = \int f_1(D|\theta)\pi(\theta|I_1) d\theta = P(D|I_1),
\end{equation}
which gives the probability of the data set $D$ given the hypothesis
$I_1$ (once $P(D|I_k)$ has been normalized to all the hypotheses). In
the same way, $Z_2$ is the probability of $D$ given the hypothesis
$I_2$. The evidences have statistical meaning. Since we can calculate
$P(D|I_1)$ and $P(D|I_2)$ we can also calculate $P(I_1|D)$ and
$P(I_2|D)$ using the Bayes' theorem obtaining the probability of a
given hypothesis given the data set $D$ and independently of the
parameters $\theta$ and $\omega$:
\begin{equation}
\label{eq:bayes_model}
P(I_m|D) = \frac{P(D|I_m)P(I_m)}{P(D)} = \frac{Z_m P(I_m)}{\sum_{l=1}^M Z_lP(I_l)},
\end{equation}
where here $M=2$ and $m = 1, 2$. The expression shown in
Eq. \ref{eq:bayes_model} is the generalization for $M$ possible
hypotheses.

Once more the prior probabilities $P(I_1)$ and $P(I_2)$ must be chosen
before the analysis. In this way, we obtain a probability mass
function in which the variables are the different hypotheses. To
compare which of the hypotheses is preferred by data, the ratio
between the posterior probabilities is performed:
\begin{equation}
\frac{P(I_1|D)}{P(I_2|D)} = \frac{Z_1}{Z_2}\frac{P(I_1)}{P(I_2)}.
\end{equation}
This ratio is called ``posterior odds'' and the ratio
$P(I_1)/P(I_2)$ is called ``prior odds''. The ratio of the
evidences $Z_1/Z_2$ is called the {\it Bayes' factor} of the
hypothesis $I_1$ over $I_2$ ($B_{1,2}$) and represents the gain of
probability of $I_1$ over the hypothesis $I_2$ after the data
analysis:
\begin{equation}
\text{posterior odds }(I_1,I_2) = B_{1,2} \times \text{prior odds }(I_1,I_2).
\end{equation}

\subsection{Predictive distributions}
Suppose that an observer wants to prepare an
experiment to infer certain parameter $\theta$ which can take values
in the $\Theta$ space with prior probabilities $\pi(\theta,I)$.  The
distribution of the random variable $X$ is given by the likelihood
function $f(x|\theta,I)$. The data distribution before the experiment is
\begin{equation}
f(\tilde{x}|I) = \int_{\Theta} f(\tilde{x}|\theta,I) \pi(\theta|I) d\theta
\end{equation}
where $\tilde{x}$ denotes unobserved data. $f(\tilde{x}|I)$ is called
the prior predictive distribution.  After the experiment has been
built and the data $D$ analysed, the knowledge about $\theta$ has
changed: $\pi(\theta,I) \rightarrow \pi(\theta|D,I)$. Now the expected
data distribution has also changed:
\begin{equation}
f(\tilde{x},I) \rightarrow f(\tilde{x}|D,I) =  
\int_{\Theta} f(\tilde{x}|\theta,I)  \pi(\theta|D,I) d\theta
\end{equation}
where $f(\tilde{x}|D,I)$ is called the posterior predictive
distribution.  This distribution can be used to compare with the
observed data distribution to get a feeling of how well the estimation
of $\theta$ fits the measured data or for future experiments.
\section{Simulations}
\label{simulations}
\begin{figure}[H]
\centering
\includegraphics[width=0.75\linewidth]{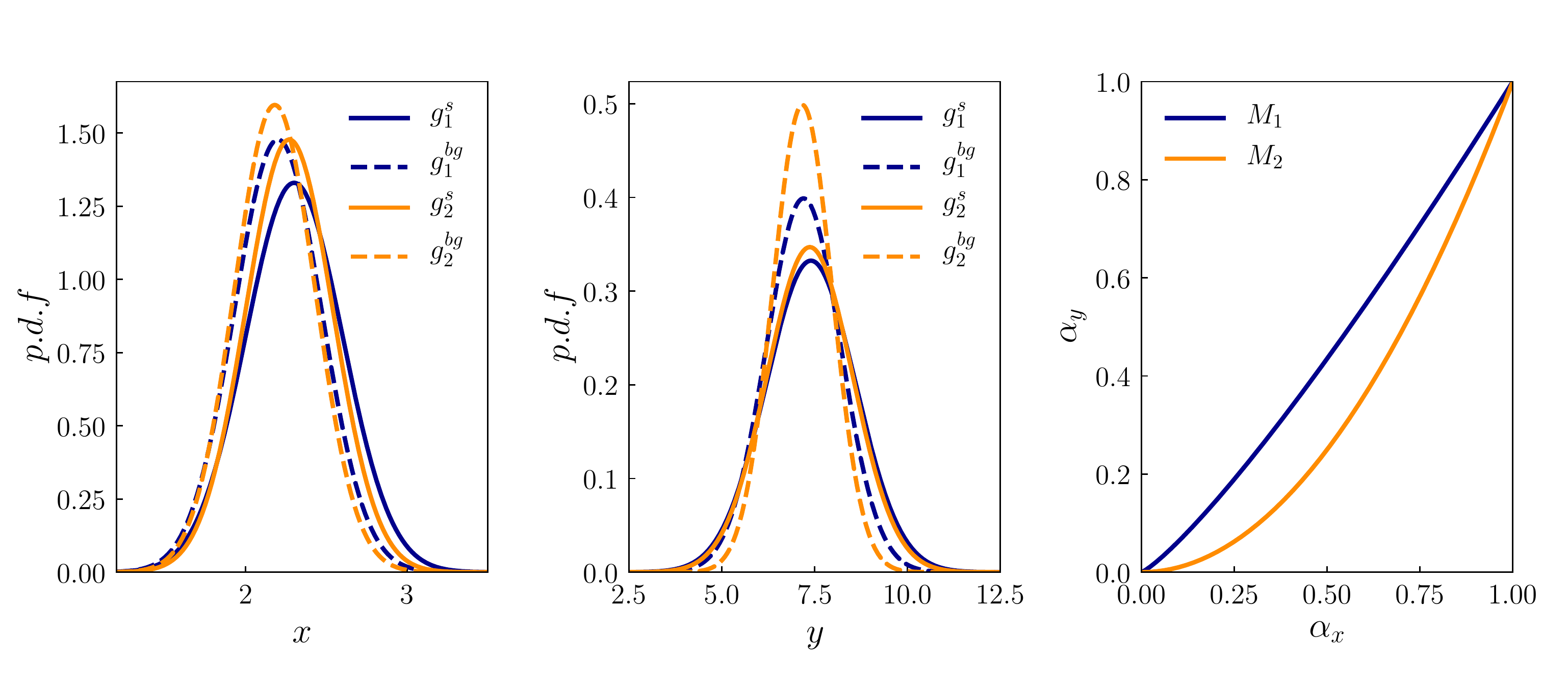}
\caption{\small{Signal and background distributions (continuous and dashed
  lines) predicted from $M_1$ (blue) and $M_2$ orange for the two
  experiments: \ex in the left panel and \ey in the center. The
  fraction of the signal in \ey as a function of the signal in \ex is
  shown in the right panel for the two models.}}
\label{fig:signals}
\end{figure}

Let \ex and \ey be two experiments measuring different observables $X$
and $Y$. The experiments are interested in to measure the fraction of
certain distribution ({\it signal}) that there is in their data. As an
example, $X$ can be the proton fraction of cosmic rays at ultra-high
energies while $Y$ can be the astrophysical photon or neutrino
fractions at energies in the PeV region. Let $M_1$ and $M_2$ two
models predicting different signals both in $X$ and $Y$ and predicting
different relations between the signals as it is illustrated in
Fig. \ref{fig:signals}. In our example $\alpha_y =
\alpha_x^{2.2}(1-\alpha_x)/2.2$ for $M_1$ and $\alpha_y =
\alpha_x^{3}(1-\alpha_x)/3$ for $M_2$.
The signal and background are normal distributions (denoted by $g^s$
and $g^{bg}$ respectively) for the two models with the following
parameters:
\begin{equation}
M_1
\begin{cases}
g_1^{s}(x) \text{ : } \mu = 2.3 \text{ and } \sigma =  0.3 \\
g_1^{bg}(x) \text{ : } \mu = 2.2 \text{ and } \sigma =  0.27 \\
g_1^{s}(y) \text{ : } \mu = 7.4 \text{ and } \sigma =  1.2 \\
g_1^{bg}(y) \text{ : } \mu = 7.2 \text{ and } \sigma =  1 \\
\end{cases}
M_2
\begin{cases}
g_2^{s}(x) \text{ : } \mu = 2.27 \text{ and } \sigma =  0.27 \\
g_2^{bg}(x) \text{ : } \mu = 2.18 \text{ and } \sigma =  0.25 \\
g_2^{s}(y) \text{ : } \mu = 7.37 \text{ and } \sigma =  1.15 \\
g_2^{bg}(y) \text{ : } \mu = 7.17 \text{ and } \sigma =  0.8 \\
\end{cases}
\end{equation}
We simulate two data samples (one for each experiment) following the
model $M_1$ with $\alpha_x^{true} = 0.3$. \ex measures 300 events and
\ey measures 200 events. For these simulations we have $\mean{x} =
2.23$ and $\sigma_x = 0.3$; $\mean{y}= 7.27$ and $\sigma_y =
1.07$. The likelihood function for the model $M_i$ ($i=1,2$) and
variable $z$ ($z=x,y$) is given by:
\begin{equation}
f(z|\alpha_z,M_i) = \alpha_zg_i^s(z) + (1-\alpha_z)g_i^{bg}(z).
\end{equation}
Now we perform two analyses: one where each experiment analyse the
data without any kind of information (Sec. \ref{sec:independent}) and
another one where the experiments use the information obtained from
the other (Sec. \ref{sec:combined}).
\subsection{Independent analyses}
\label{sec:independent}
In this approach the experiments have no any prior information but
they are interested in the fraction of the signal, then the fraction
of signal plus the fraction of background must be one. For this
reason, each experiment choose a uniform distribution between $0$ and
$1$ as its prior.
\begin{figure}[H]
\centering
\begin{subfigure}[t]{0.3\linewidth}
\centering
\includegraphics[width=1\linewidth]{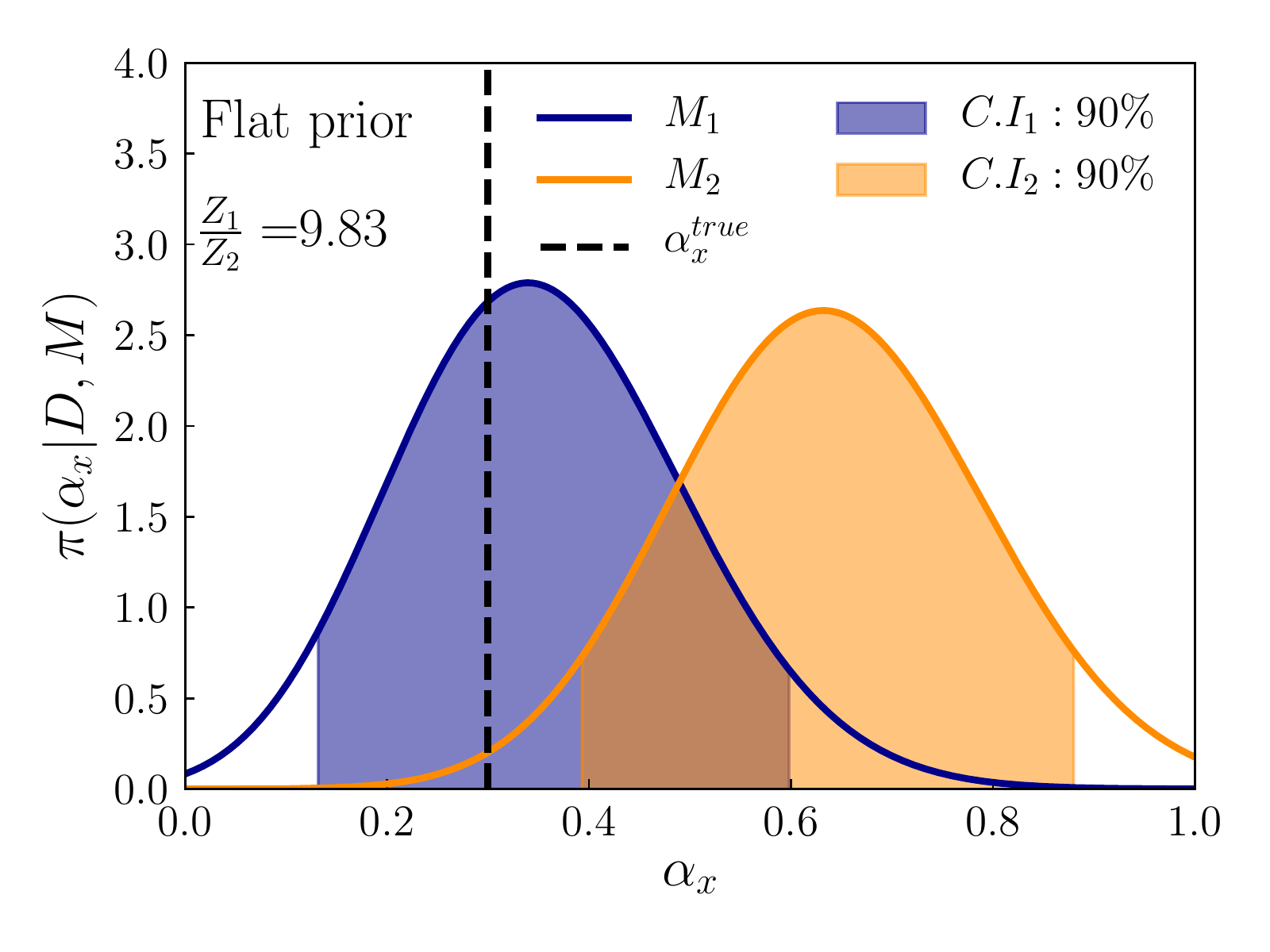}
\caption{}
\end{subfigure}
\begin{subfigure}[t]{0.3\linewidth}
\centering
\includegraphics[width=1\linewidth]{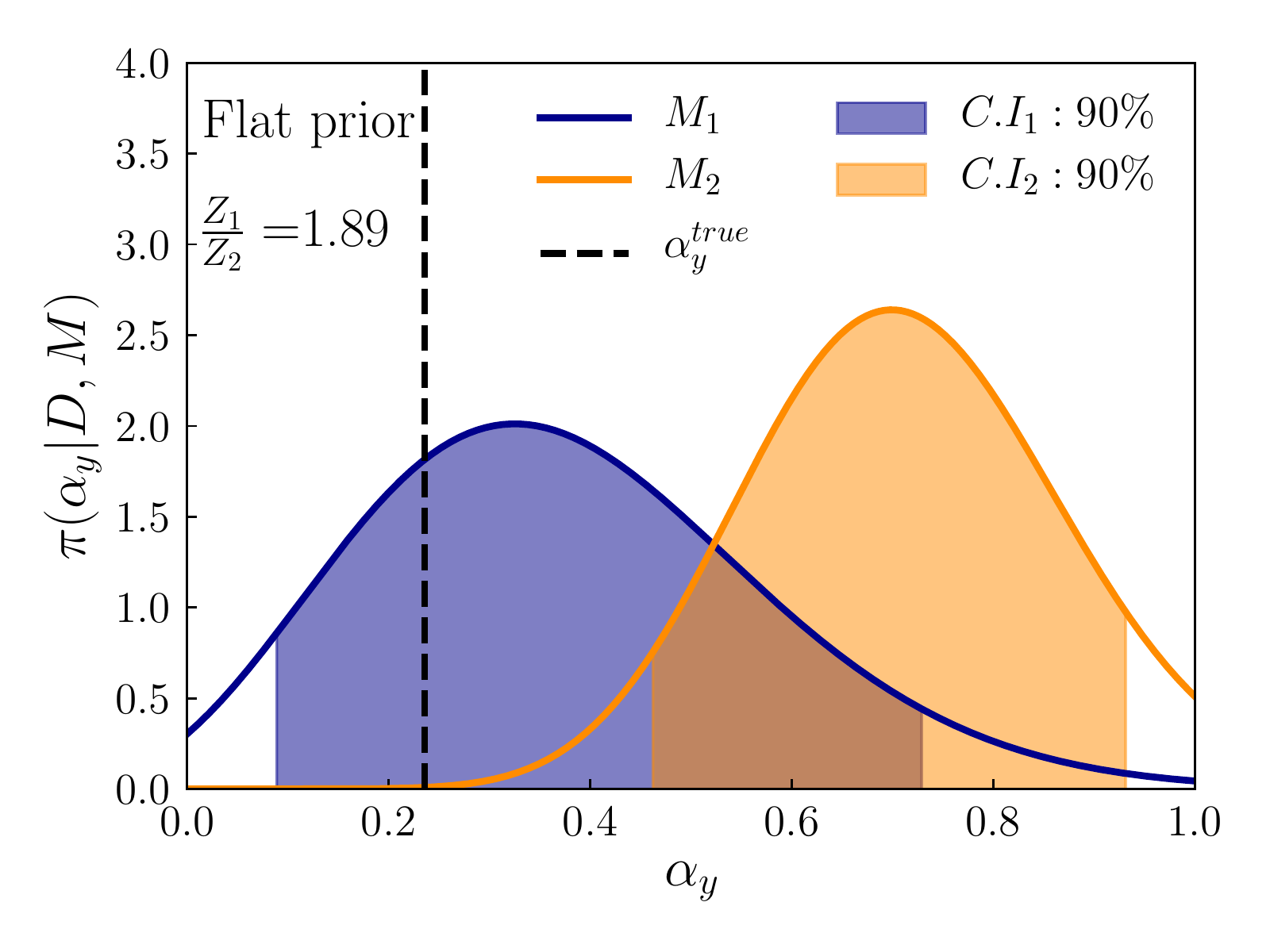}
\caption{}
\end{subfigure}
\caption{\small{Posterior probability distributions of \ex (a) and \ey
    (b) for the fraction of the interesting signal assuming the
    different models: $M_1$ (blue) and $M_2$ (orange).}}
\label{fig:posterior-flats}
\end{figure}
In Fig. \ref{fig:posterior-flats} the posterior probabiltiy
distributions of the signals for each experiment under the assumption
of the different models are displayed. We obtain numerically that \ex
obtains $\mean{\alpha}_x = 0.33$ and a C.I at $90\%$ $[0.13,0.6]$
assuming $M_1$ while assuming $M_1$ \ex obtains $\mean{\alpha}_x =
0.6$ and $[0.4,0.9]$ as the posterior mean value of the fraction of
the signal and C.I respectively.  \ey obtains $\mean{\alpha}_y = 0.38$
and C.I $0.09,0.74$ assuming $M_1$ and $\mean{\alpha}_y = 0.62$ as
fraction of the interesting signal with $[0.46,0.93]$ as a C.I
assuming $M_2$. Since each experiment assumes $P(M_1) = P(M_2) = 0.5$,
before the analysis, \ex arrives to the conclusion that $M_1$ is
almost ten times most probable than $M_2$ while the resolution to
discriminate between the models in \ey is smaller and for this
experiment $P(M_1|D)/P(M_1|D) \sim 2$.
\subsection{Combined analyses}
\label{sec:combined}
\begin{figure}[H]
\centering
\begin{subfigure}[t]{0.32\linewidth}
\centering
\includegraphics[width=1\linewidth]{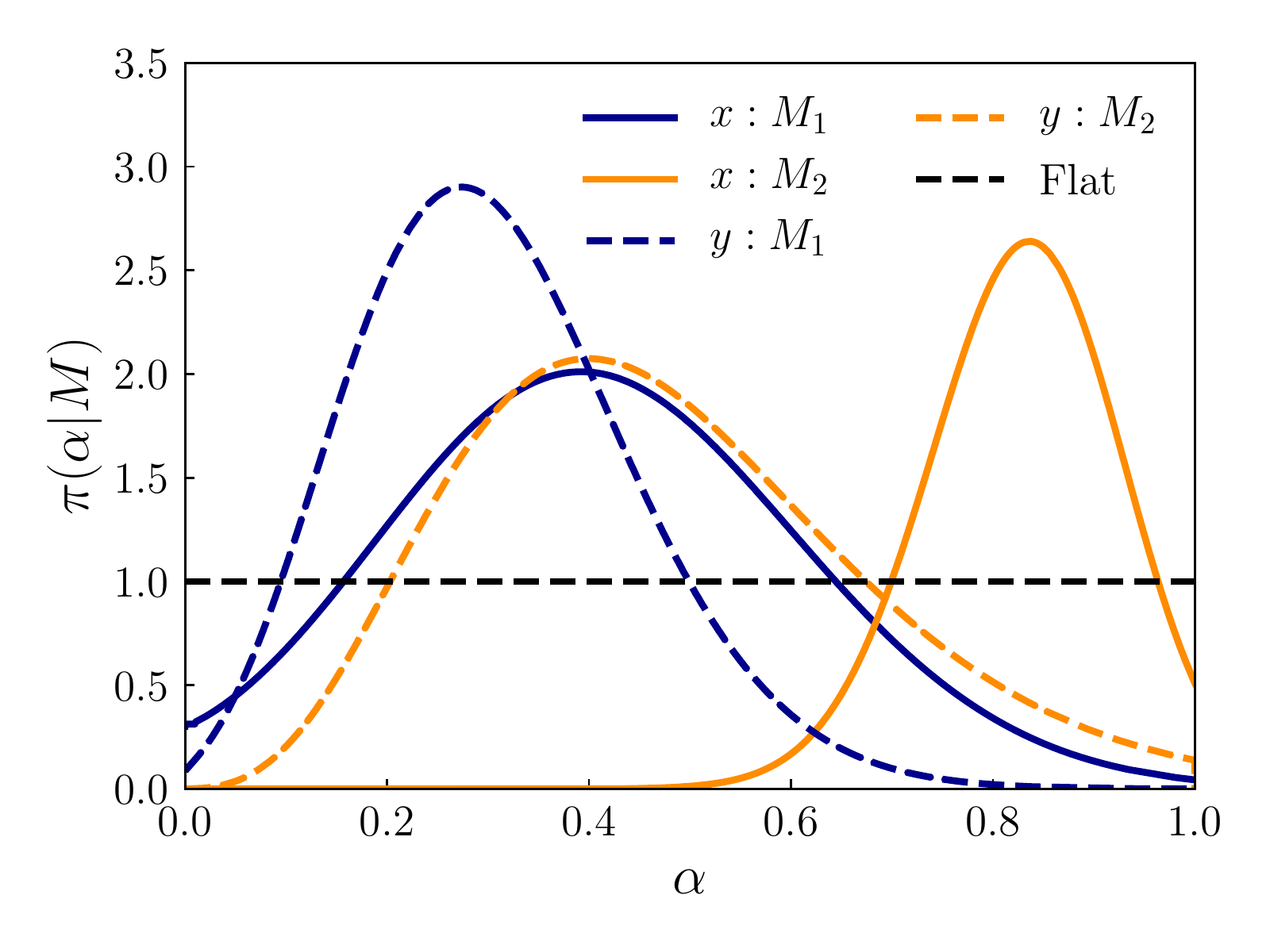}
\caption{}
\end{subfigure}
\begin{subfigure}[t]{0.32\linewidth}
\centering
\includegraphics[width=1\linewidth]{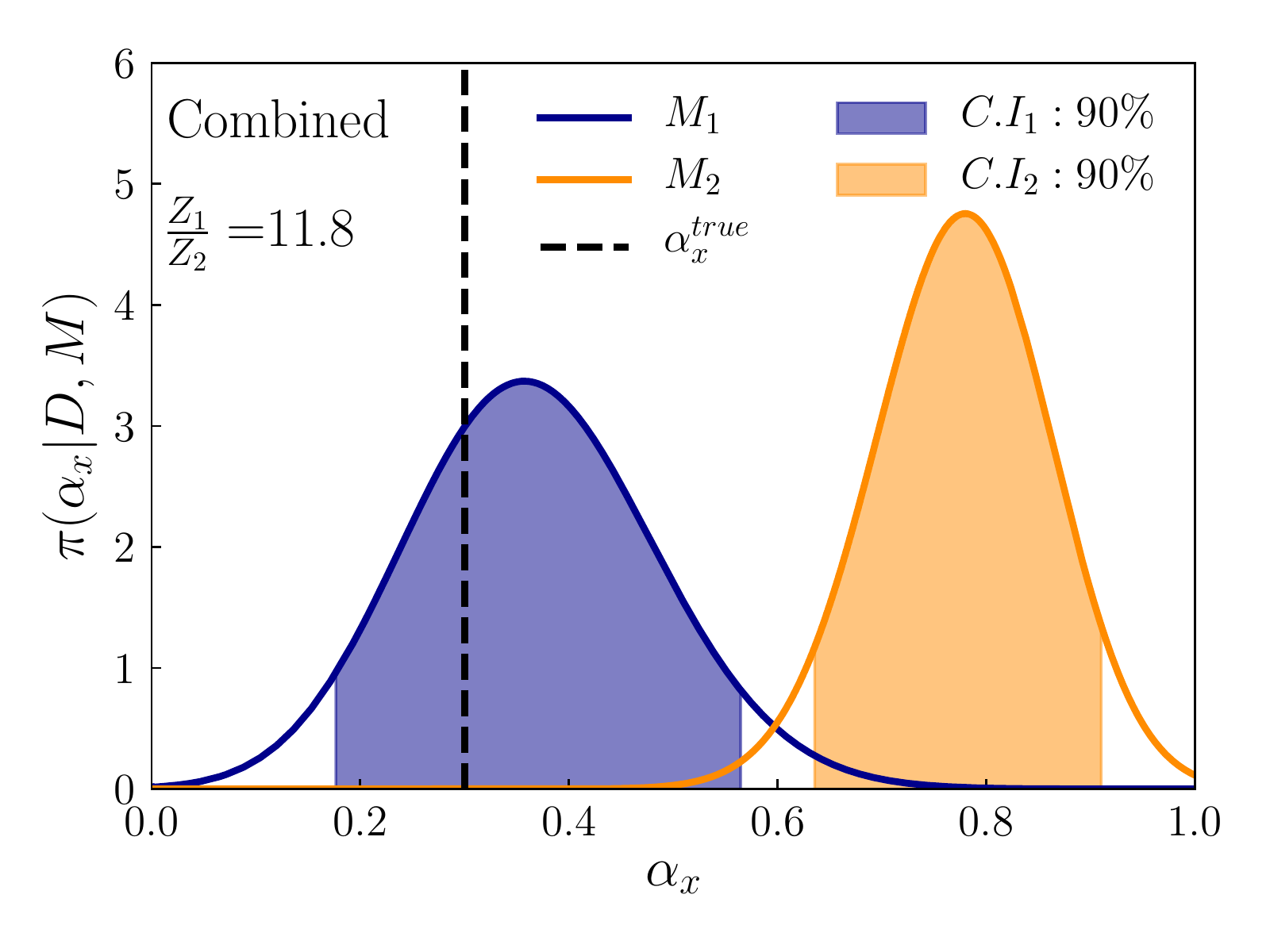}
\caption{}
\end{subfigure}
\begin{subfigure}[t]{0.32\linewidth}
\centering
\includegraphics[width=1\linewidth]{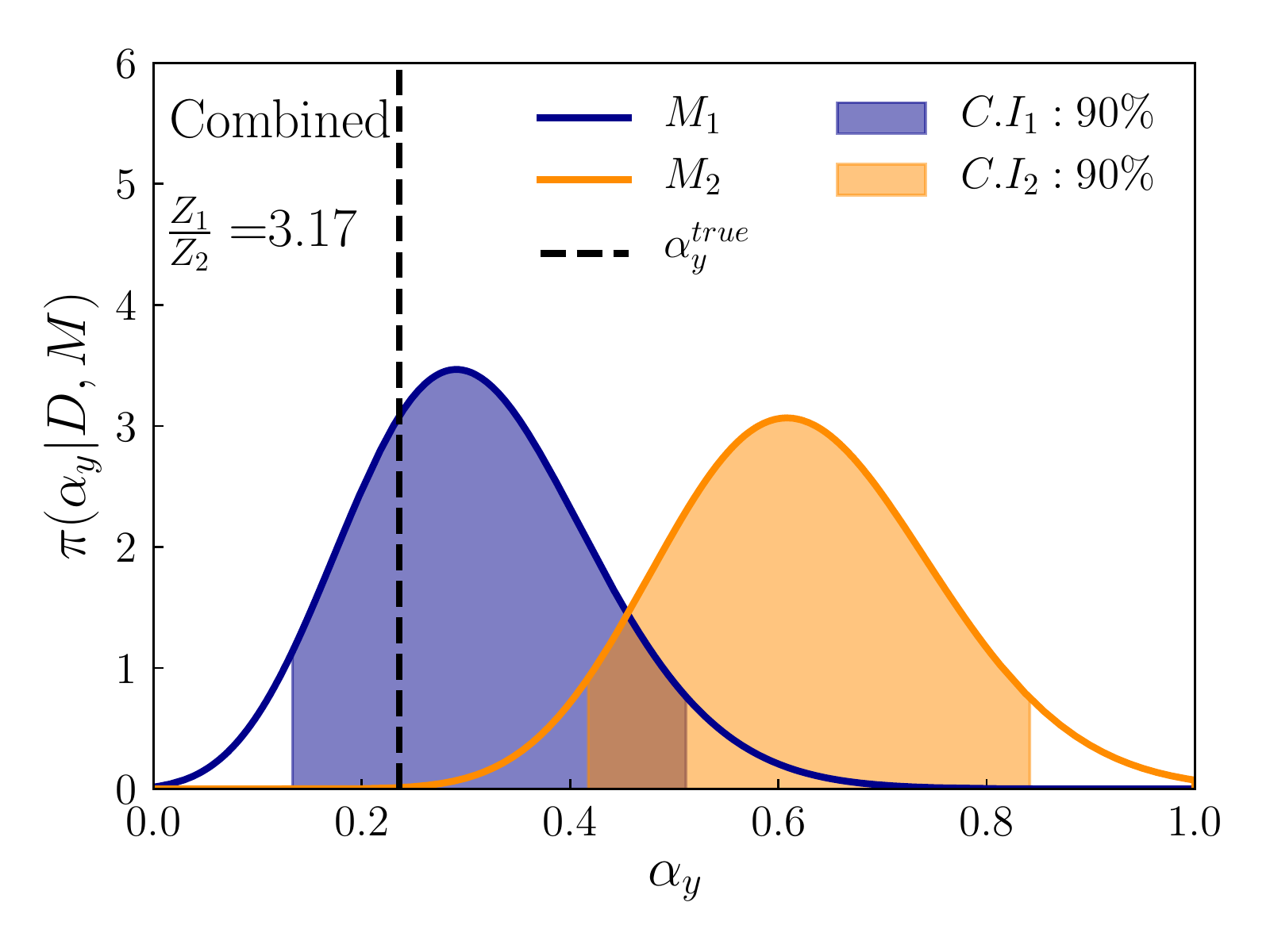}
\caption{}
\end{subfigure}
\caption{\small{Prior probabilities for the signals given the
    independent analysis in panel (a): prior for $\alpha_x$ given the
    results of \ey assuming $M_1$ ($M_2$) is shown as continuous blue
    (orange) line; prior for $\alpha_y$ given the results of \ex
    assuming $M_1$ ($M_2$) are shown as dashed blue (orange)
    line. They are compared with the uniform prior taken in the
    independent analysis (black dashed line). The posterior
    distributions for each experiment are shown in panels (b) and (c)
    for \ex and \ey respectively.}}
\label{fig:posterior-nonflats}
\end{figure}

When one experiment has analysed some data, its prior knowledge
change, and these change can be use for the same experiment to analyse
new data or for another experiment. In this example we show how the
results of each experiment is used by the other. Assuming the results
of \ex in the previous section \ey can modify the prior of $\alpha_y$
for each theoretical model or scenario. In the same way, \ex can do
the same in sight of the analysis done by \ey. These new priors are
shown together with the new results in
Fig. \ref{fig:posterior-nonflats}.
One can observe that by including the results of one experiment in the
other the results change. Now \ex obtains that the posterior odds in
favour of $M_1$ are: $P(M_1|D)/P(M_2|D) = 11.8 \times 1.89 \approx
22$, increasing the evidence in favour of the model $M_1$. When \ey
analyse its data taking into account the results of \ex the posterior
odds also increase being now $P(M_1|D)/P(M_2|D) \approx 31$. Therefore
both experiments have reasons to beleave that the true model is $M_1$
and the joined results will be $\mean{\alpha_x} = 0.36$ and
$\mean{\alpha_y} = 0.31$ with C.I $[0.18,0.56]$ and $[0.13,0.51]$
respectively being $M_1$ at least $22$ times more probable than $M_2$.

\begin{figure}[H]
\centering
\begin{subfigure}[t]{0.4\linewidth}
\centering
\includegraphics[width=1\linewidth]{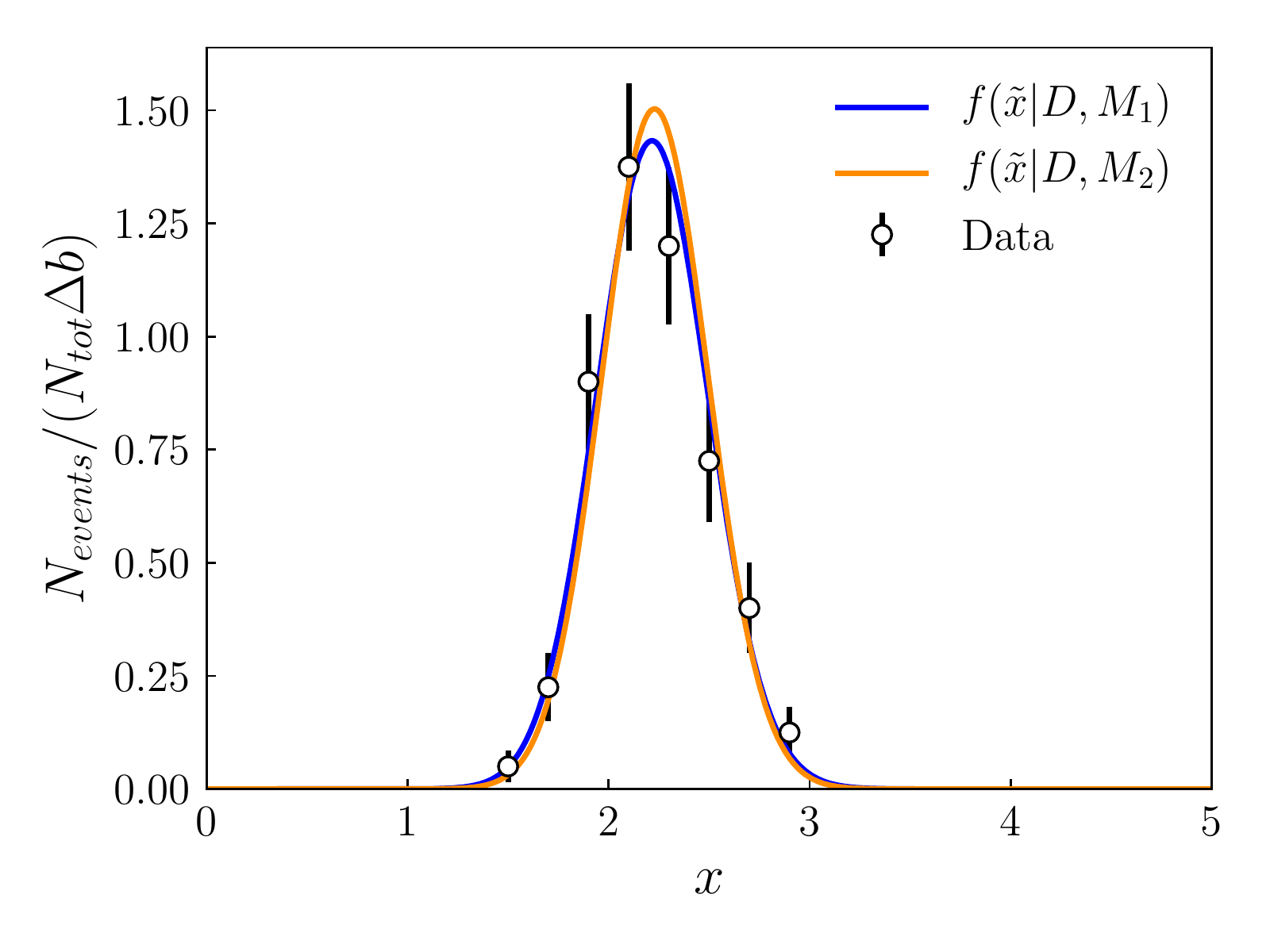}
\caption{}
\end{subfigure}
\begin{subfigure}[t]{0.4\linewidth}
\centering
\includegraphics[width=1\linewidth]{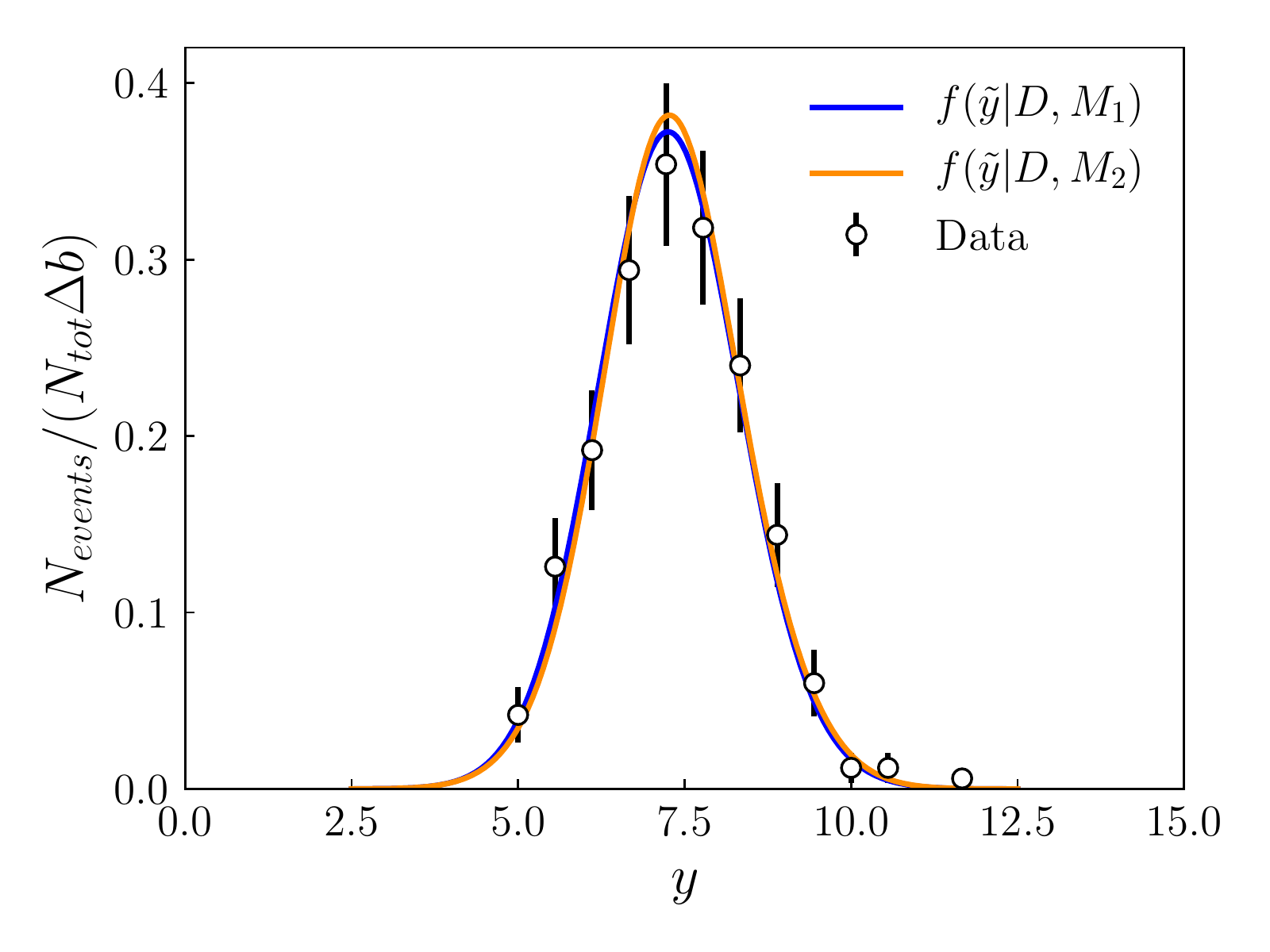}
\caption{}
\end{subfigure}
\caption{\small{Posterior predictive distributions for \ex (a) and \ey
    (b) compared with the observed data.}}
\label{fig:predictive_combined}
\end{figure}

Finally, the posterior predictive distributions taking the results of
the combined analysis are shown in Fig. \ref{fig:predictive_combined}.
Even though the data can be well described by the two models, the
Bayesian combined analysis permits us distinguish numerically between
the two models.
\section{Conclusions}
\label{conclusions}
The Bayesian approach for the combination of different measurements
and detectors has been presented and tested using simulations. With
these methods the estimation of the parameters of interest and the
discrimination among different theoretical models or scenarios can be
improved using past or present experimental results from different
experiments.

In this work we show how the combination of the information obtained
with two different detectors can improve the parameter estimation,
reduce the uncertainty and distinguish between theoretical models that
can explain the same data.

\end{document}